# Computing with and without arbitrary large numbers

(Extended abstract)

Michael Brand
michael.brand@alumni.weizmann.ac.il

Faculty of IT, Monash University
Clayton, VIC 3800
Australia

September 12, 2018

**Abstract**

In the study of random access machines (RAMs) it has been shown that the availability of an extra input integer, having no special properties other than being sufficiently large, is enough to reduce the computational complexity of some problems. However, this has only been shown so far for specific problems. We provide a characterization of the power of such extra inputs for general problems.

To do so, we first correct a classical result by Simon and Szegedy (1992) as well as one by Simon (1981). In the former we show mistakes in the proof and correct these by an entirely new construction, with no great change to the results. In the latter, the original proof direction stands with only minor modifications, but the new results are far stronger than those of Simon (1981).

In both cases, the new constructions provide the theoretical tools required to characterize the power of arbitrary large numbers.

## 1 Introduction

The Turing machine (TM), first introduced in [17], is undoubtedly the most familiar computational model. However, for algorithm analysis it often fails to adequately represent real-life complexities, for which reason the random access machine (RAM), closely resembling the intuitive notion of an idealized computer, has become the common choice in algorithm design. Ben-Amram and Galil [2] write "The RAM is intended to model what we are used to in conventional programming, idealized in order to be better accessible for theoretical study."

Here, "what we are used to in conventional programming" refers, among other things, to the ability to manipulate high-level objects by basic commands. However, this ability comes with some unexpected side effects. For example, one can consider a RAM that takes as an extra input an integer that has no special property other than being "large enough". Contrary to intuition, it has been shown that such arbitrary large numbers (ALNs) can lower problem time complexities. For example, [4] shows that the availability of ALNs lowers the arithmetic time complexity[1] of calculating $2^{2^x}$ from $\Theta(x)$ to $\Theta(\sqrt{x})$. However, all previous attempts to characterize the contribution of ALNs dealt with problem-specific methods of exploiting such inputs, whereas the present work gives, for the first time, a broad characterization of the scenarios in which arbitrary numbers do and those in which they do not increase computational power.

In order to present our results, we first redefine, briefly, the RAM model. (See [1] for a more formal introduction.)

Computations on RAMs are described by *programs*. RAM programs are sets of *commands*, each given a *label*. Without loss of generality, labels are taken to be consecutive integers. The bulk of RAM commands belong to one of two types. One type is an *assignment*. It is described by a triplet containing a $k$-ary operator, $k$ operands and a target. The other type is a *comparison*. It is given two operands and a comparison operator, and is equipped with labels to proceed to if the comparison is evaluated as either true or false. Other command-types include unconditional jumps and execution halt commands.

The execution model for RAM programs is as follows. The RAM is considered to have access to an infinite set of *registers*, each marked by a non-negative integer. The input to the program is given as the initial state of the first registers. The rest of the registers are initialized to 0. Program execution begins with the command labeled 1 and proceeds sequentially, except in comparisons (where execution proceeds according to the result of the comparison) and in jumps. When executing assignments, the $k$-ary operator is evaluated based on the values of the $k$ operands and the result is placed in the target register. The output of the program is the state of the first registers at program termination.

In order to discuss the computational power of RAMs, we consider only RAMs that are comparable in their input and output types to TMs. Namely, these will be the RAMs whose inputs and outputs both lie entirely in their first register. We compare these to TMs working on one-sided-infinite tapes over a binary alphabet, where "0" doubles as the blank. A RAM will be considered equivalent to a TM if, given as an input an integer whose binary encoding is the initial state of the TM's tape, the RAM halts with a non-zero output value if and only if the TM accepts on the input.

Furthermore, we assume, following e.g. [6], that all explicit constants used as operands in RAM programs belong to the set $\{0, 1\}$. This assumption does not make a material difference to the results, but it simplifies the presentation.

In this paper we deal with RAMs that use non-negative integers as their register contents. This is by far the most common choice. A RAM will be indicated by RAM[$op$], where $op$ is the set of basic operations supported by the RAM. These basic operations are assumed to execute in a single unit of time. We use the syntax $f(n)$-RAM[$op$] to denote the set of problems solvable in $f(n)$ time by a RAM[$op$], where $n$ is the bit-length of the input. Replacing "RAM[$op$]" by "TM" indicates that the computational model used is a Turing machine.

---

[1]Arithmetic complexity is the computational complexity of a problem under the RAM[$+, \overset{.}{-}, \times, \div$] model, which is defined later on in this section.

Note that because registers only store non-negative integers, such operations as subtraction cannot be supported without tweaking. The customary solution is to replace subtraction by "natural subtraction", denoted "$\dotdiv$" and defined by $a \dotdiv b \stackrel{\text{def}}{=} \max(a - b, 0)$. We note that if the comparison operator "$\leq$" (testing whether the first operand is less than or equal to the second operand) is not supported by the RAM directly, the comparison "$a \leq b$" can be simulated by the equivalent equality test "$a \dotdiv b = 0$". Testing for equality is always assumed to be supported.

By the same token, regular bitwise negation is not allowed, and $\neg a$ is tweaked to mean that the bits of $a$ are negated only up to and including its most significant "1" bit.

Operands to each operation can be explicit integer constants, the contents of explicitly named registers or the contents of registers whose numbers are specified by other registers. This last mode, which can also be used to define the target register, is known as "indirect addressing". In [3] it is proved that for the RAMs considered here indirect addressing has no effect. We therefore assume throughout that it is unavailable to the RAMs.

The following are two classical results regarding RAMs. Operations appearing in brackets within the operation list are optional, in the sense that the theorem holds both when the operation is part of $op$ and when it is not.

**Theorem 1** ([14]). *PTIME-RAM$[+, [\dotdiv], [\times], \leftarrow, [\rightarrow], \text{Bool}] = \text{PSPACE}$*

and

**Theorem 2** ([13]). *PTIME-RAM$[+, \dotdiv, /, \leftarrow, \text{Bool}; \leq] = \text{ER}$, where ER is the set of problems solvable by Turing machines in*

$$\left.2^{2^{\cdot^{\cdot^{\cdot^{2}}}}}\right\}n$$

*time, where $n$ is the length of the input.*

Here, "/" indicates exact division, which is the same as integer division (denoted "$\div$") but is only defined when the two operands divide exactly. The operations "$\leftarrow$" and "$\rightarrow$" indicate left shift ($a \leftarrow b = a \times 2^b$) and right shift ($a \rightarrow b = a \div 2^b$), and *Bool* is shorthand for the set of all bitwise Boolean functions.

In this paper, we show that while Theorem 1 is correct, its original proof is not. Theorem 2, on the other hand, despite being a classic result and one sometimes quoted verbatim (see, e.g., [16]), is, in fact, erroneous.

We re-prove the former here, and replace the latter by a stronger result, for the introduction of which we first require several definitions.

**Definition 1** (Expansion Limit). Let $M = M_{op}(t, inp)$ be the largest number that can appear in any register of a RAM$[op]$ working on $inp$ as its input, during the course of its first $t$ execution steps.

We define $\text{EL}_{op}(f(n))$ to be the maximum of $M_{op}(f(n), inp)$ over all values of $inp$ for which $\text{len}(inp) \leq n$. This is the maximum number that can appear in any register of a RAM$[op]$ that was initialized by an input of length at most $n$, after $f(n)$ execution steps.

The subscript $op$ may be omitted if understood from the context.

As a slight abuse of notation, we use EL($t$) to be the maximum of $M_{op}(t, inp)$ over all $inp$ of length at most $n$, when $n$ is understood from the context and $t$ is independent of $n$. (The following definition exemplifies this.)

**Definition 2** (RAM-Constructability)**.** A set of operations $op$ is *RAM-constructable* if the following two conditions are satisfied: (1) there exists a RAM program that, given $inp$ and $t$ as its inputs, with $n$ being the length of $inp$, returns in O($t$) time a value no smaller than $\mathrm{EL}_{op}(t)$, and (2) each operation in $op$ is computable in EL(O($l$)) space on a Turing machine, where $l$ is the total length of all operands and of the result.

Our results are as follows.

**Theorem 3.** *For a RAM-constructable $op \supseteq \{+, /, \leftarrow, Bool\}$ and any function $f(n)$,*

$$\begin{aligned} O(f(n))\text{-}RAM[op] &= EL_{op}(O(f(n)))\text{-}TM \\ &= N\text{-}EL_{op}(O(f(n)))\text{-}TM \\ &= EL_{op}(O(f(n)))\text{-}SPACE\text{-}TM \\ &= N\text{-}EL_{op}(O(f(n)))\text{-}SPACE\text{-}TM, \end{aligned}$$

*where the new notations refer to nondeterministic Turing machines, space-bounded Turing machines and nondeterministic space-bounded Turing machines, respectively.*

Among other things, this result implies for polynomial-time RAMs that their computational power is far greater than ER, as was previously believed.

The theoretical tools built for proving Theorem 3 and re-proving Theorem 1 then allow us to present the following new results regarding the power of arbitrary large numbers.

**Theorem 4.** *PTIME-ARAM[$+, [\dot{-}], [\times], \leftarrow, [\rightarrow], Bool$] = PSPACE*

**Theorem 5.** *Any recursively enumerable (r.e.) set can be recognized by an ARAM[$+, /, \leftarrow, Bool$] in O(1) time.*

Here, "ARAM" is the RAM model assisted by an arbitrary large number. Formally, we say that a set $S$ is computable by an ARAM[$op$] in $f(n)$ time if there exists a Boolean function $g(inp, x)$, computable in $f(n)$ time on a RAM[$op$], such that $inp \in S$ implies $g(inp, x) \neq 0$ for almost all $x$ (all but a finite number of $x$) whereas $inp \notin S$ implies $g(inp, x) = 0$ for almost all $x$. Here, $n$ conventionally denotes the bit length of the input, but other metrics are also applicable.

We see, therefore, that the availability of arbitrary numbers has no effect on the computational power of a RAM without division. However, for a RAM equipped with integer division, the boost in power is considerable, to the extent that any problem solvable by a Turing machine in any amount of time or space can be solved by an ARAM in O(1) time.

## 2 Models without division

### 2.1 Errata on [14]

We begin with a definition.

**Definition 3** (Straight Line Program). A *Straight Line Program*, or SLP[*op*], is a list of tuples, $s_2, \ldots, s_n$, where each $s_i$ is composed of an operator, $s_i^{op} \in op$, and $k$ integers, $s_i^1, \ldots, s_i^k$, all in the range $0 \leq s_i^j < i$, where $k$ is the number of operands taken by $s_i^{op}$. This list is to be interpreted as a set of computations, whose targets are $v_0, \ldots, v_n$, which are calculated as follows: $v_0 = 0$, $v_1 = 1$, and for each $i > 1$, $v_i$ is the result of evaluating the operator $s_i^{op}$ on the inputs $v_{s_i^1}, \ldots, v_{s_i^k}$. The output of an SLP is the value of $v_n$.

A technique first formulated in a general form in [11] allows results on SLPs to be generalized to RAMs. Schönhage's theorem, as worded for the special case that interests us, is that if there exists a Turing machine, running on a polynomial-sized tape and in finite time, that takes an SLP[*op*] as input and halts in an accepting state if and only if $v_n$ is nonzero, then there also exists a TM running on a polynomial-sized tape that simulates a RAM[*op*]. This technique is used both in [14] and in our new proof.

The proof of [14] follows this scheme, and attempts to create such a Turing machine. In doing so, this TM stores monomial-based representations of certain powers of two. These are referred to by the paper as "monomials" but are, for our purposes, integers.

The main error in [14] begins with the definition of a relation, called "vicinity", between monomials, which is formulated as follows.

> We define an equivalence relation called *vicinity* between monomials. Let $M_1$ and $M_2$ be two monomials. Let $B$ be a given parameter. If $M_1/M_2 < 2^{2^B}$ [...], then $M_1$ is in the vicinity of $M_2$. The symmetric and transitive closure of this relation gives us the full vicinity relation. As it is an equivalence relation, we can talk about two monomials being in the same vicinity (in the same equivalence class).

It is unclear from the text whether the authors' original intention was to define this relation in a universal sense, as it applies to the set of all monomials (essentially, the set of all powers of two), or whether it is only defined over the set of monomials actually used by any given program. If the former is correct, any two monomials are necessarily in the same vicinity, because one can bridge the gap between them by monomials that are only a single order of magnitude apart. If the latter is correct, it is less clear what the final result is. The paper does not argue any claim that would characterize the symmetric and transitive closure in this case.

However, the paper does implicitly assume throughout that the vicinity relation, as originally defined (in the $M_1/M_2 < 2^{2^B}$ sense) is *its own* symmetric and transitive closure. This is used in the analysis by assuming for any $M_i$ and $M_j$ which are in the same vicinity (in the equivalence relation sense) that they also satisfy $2^{-(2^B)} < M_i/M_j < 2^{2^B}$, i.e. they are in the same vicinity also in the restrictive sense.

Unfortunately, this claim is untrue. It is quite possible to construct an SLP that violates this assumption, and because the assumption is central to the entire algorithm, the proof does not hold.

We therefore provide here an alternate algorithm, significantly different from the original, that bypasses the entire "vicinity" issue.

## 2.2 Our new construction

Our proof adapts techniques from two previous papers: [5] (which uses lazy evaluation to perform computations on operands that are too long to fit into a polynomial-sized tape) and [12] (which stores operands in a hierarchical format that notes only the positions of "interesting bits", these being bit positions whose values are different than those of the less significant bit directly preceding them). The former method is able to handle multiplication but not bit shifting and the latter the reverse. We prove Theorem 1 using a sequence of lemmas. All algorithms described are available as C++ code in Appendix A.

**Lemma 1.** *In an $SLP[+, \dot{-}, \times, \leftarrow, \rightarrow, Bool]$, the number of interesting bits in the output $v_n$ grows at most exponentially with $n$. There exists a Turing machine working in polynomial space that takes such an SLP as its input, and that outputs an exponential-sized set of descriptions of bit positions, where bit positions are described as functions of $v_0, \ldots, v_{n-1}$, such that the set is a superset of the interesting bit positions of $v_n$.*

The fact that the number of interesting bits grows only exponentially given this operation set was noted in [14]. Our proof follows the reasoning of the original paper.

*Proof.* Consider, for simplicity, the instruction set $op = \{+, \times, \leftarrow\}$. Suppose that we were to change the meaning of the operator "$\leftarrow$", so that, instead of calculating $a \leftarrow b = a \times 2^b$, its result would be $a \leftarrow b = aX$, where $X$ is a formal parameter, and a new formal parameter is generated every time the "$\leftarrow$" operator is used. The end result of the calculation will now no longer be an integer but rather a polynomial in the formal parameters. The following are some observations regarding this polynomial.

1. The number of formal parameters is at most $n$, the length of the SLP.

2. The power of each formal parameter is at most $2^{n-k}$, where $k$ is the step number in which the parameter was defined. (This exponent is at most doubled at each step in the SLP. Doubling may happen, for example, if the parameter is multiplied by itself.)

3. The sum of all multiplicative coefficients in the polynomial is at most $2^{2^{n-2}}$. (During multiplication, the sum of the product polynomial's coefficients is the product of the sums of the operands' coefficients. As such, this value can at most square itself at each operation. The maximal value it can attain at step 2 is 2.)

If we were to take each formal variable, $X$, that was created at an "$a \leftarrow b$" operation, and substitute in it the value $2^b$ (a substitution that [14] refers to as the "standard evaluation"), then the value of the polynomial will equal the value of the SLP's output. We claim that if $p$ is an interesting bit position, then there is some product of formal variables appearing as a monomial in the result polynomial such that its standard evaluation is $2^x$, and $p \geq x \geq p - 2^n$.

The claim is clearly true for $n = 0$ and $n = 1$. For $n > 1$, we will make the stronger claim $p \geq x \geq p - 2^{n-2} - 2$. To prove this, note that any monomial whose standard evaluation is greater than $2^p$ cannot influence the value of bit $p$ and cannot make it "interesting". On the other hand, if all remaining monomials are smaller than $p - 2^{n-2} - 2$, the total value that they carry within the polynomial is smaller than $2^{p-2^{n-2}-2}$ times the sum of their coefficients, hence smaller than $2^{p-2}$. Bits $p-1$ and $p$, however, are both zero. Therefore, $p$ is not an interesting bit.

We proved the claim for the restricted operation set $\{+,\times,\leftarrow\}$. Adding logical AND ("$\wedge$") and logical OR ("$\vee$") can clearly not change the fact that bits $p-1$ and $p$ are both zero, nor can it make the polynomial coefficients larger than $2^{2^{n-2}}$.

Incorporating "$\stackrel{.}{-}$" and "$\neg$" into the operation set has a more interesting effect: the values of bit $p-1$ and $p$ can both become "1". This will still not make bit $p$ interesting, but it does require a small change in the argument. Instead of considering polynomials whose coefficients are between 0 and $2^{2^{n-2}}$, we can now consider polynomials whose coefficients are between $-2^{2^{n-2}}$ and $2^{2^{n-2}}$. This changes the original argument only slightly, in that we now need to argue that in taking the product over two polynomials the sum of the absolute values of the coefficients of the product is no greater than the product of the sums of the absolute values of the coefficients of the operands.

Similarly, adding "$\rightarrow$" into consideration, we no longer consider only formal variables of the form $a \leftarrow b = aX$ but also $a \rightarrow b = \lfloor aY \rfloor$, where the standard evaluation of $Y$ is $2^{-b}$ and $\lfloor \cdot \rfloor$ is treated as a bitwise Boolean operation (in the sense that, conceptually, it zeroes all bit positions that are "to the right of the decimal point" in the product).

We can therefore index the set of interesting bits by use of a tuple, as follows. If $i_1,\ldots,i_k$ are the set of steps for which $s_{i_j}^{op} \in \{\leftarrow,\rightarrow\}$, the tuple will contain one number between 0 and $2^{n-i_j}$ for each $1 \leq j \leq k$, to indicate the exponent of the formal parameter added at step $i_j$, and an additional $k+1$'th element, between 0 and $2^n$ to indicate a bit offset from this bit position.

Though this tuple may contain many non-interesting bits, or may describe a single bit position by many names, it is a description of a super-set of the interesting bits in polynomial space. □

In Appendix A, such an enumeration is implemented by the method `Index::next`. We refer to the set of bit positions thus described as the *potentially-interesting* bits, or *po-bits*, of the SLP.

**Lemma 2.** *Let $\mathcal{S},\mathcal{S}' \in SLP[+,\stackrel{.}{-},\times,\leftarrow,\rightarrow,Bool]$. Let $\mathcal{O}$ be an Oracle that takes $\mathcal{S}'$ as input and outputs the descriptions of all its po-bits in order, from least-significant to most-significant, without repetitions. There exists a TM working in polynomial space but with access to $\mathcal{O}$ that takes as inputs $\mathcal{S}$ and the description of a po-bit position, $i$, of $\mathcal{S}$, and that outputs the $i$'th bit of the output of $\mathcal{S}$.*

*Proof.* Given a way to iterate over the po-bits in order, the standard algorithms for most operations required work as expected. For example, addition can be performed bit-by-bit if the bits of the operands are not stored, but are, rather, calculated recursively whenever they are needed. The depth of the recursion required in this case is at most $n$. (See `Add::eval` in Appendix A.)

The fact that iterating only over the po-bits, instead of over all bit positions, makes no difference to the results is exemplified in Figure 1.

As can be seen, not only are the non-po-bits all equal to the last po-bit preceding them, in addition, the carry bit going over from the last po-bit to the first non-po-bit is the same as the carry bit carried over from the last non-po-bit to the first po-bit. Because of this, the sequential carry bits across non-po-bits (depicted in light blue in Figure 1) can be replaced by a single non-contiguous carry operation (the red arrow).

This logic works just as well for subtraction and Boolean operations. The only operation acting differently is multiplication. Implementing multiplication directly leads to incorrect results. Instead, we re-encode the operand bits in a way that reflects our original observation, that the operands can be taken to be polynomials with small coefficients in absolute value, though these coefficients may not necessarily be nonnegative.

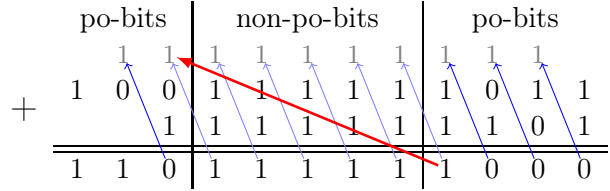

Figure 1: An example of summing two numbers

The new encoding is as follows: going from least significant bit to most significant bit, a "0" bit is encoded as a 1 if preceded by a "1" and as 0, otherwise. A "1" bit is encoded as a 0 if preceded by a "1" and as $-1$, otherwise. It is easy to see that a number, $A$, encoded in regular binary notation but including a leading zero by a $\{0, 1\}$ sequence, $a_0, \ldots, a_k$, denoting coefficients of a power series $A = \sum_{i=0}^{k} a_i 2^i$, does not change its value if the $a_i$ are switched for the $b_i$ that are the result of the re-encoding procedure described. The main difference is that now the value of all non-po-bits is 0. (See `Mult::eval` in Appendix A.)

Proving that multiplication works correctly after re-encoding is done by observing its base cases and bilinear properties. The carry in the calculation is exponential in size, so can be stored using a polynomial number of bits. □

**Lemma 3.** *Let $\mathcal{Q}$ be an Oracle that takes an $\mathcal{S} \in SLP[+, \dot{-}, \times, \leftarrow, \rightarrow, Bool]$ and two po-bit positions of $\mathcal{S}$ and determines which position is the more significant. Given access to $\mathcal{Q}$, Oracle $\mathcal{O}$, described in Lemma 2, can be implemented as a polynomial space Turing machine.*

*Proof.* Given an Oracle able to compare between indices, the ability to enumerate over the indices in an arbitrary order allows creation of an ordered enumeration. Essentially, we begin by choosing the smallest value, then continue sequentially by choosing, at each iteration, the smallest value that is still greater than the current value. This value is found by iterating over all index values in an arbitrary order and trying each in turn. □

In Appendix A, this algorithm is implemented by the method `Index::operator++`.

**Lemma 4.** *Oracle $\mathcal{Q}$, described in Lemma 3, can be implemented as a polynomial space Turing machine.*

*Proof.* Recall that an index position is an affine function of the coefficients of the formal variables introduced, in their standard evaluations. To determine which of two indices is larger, we subtract these, again reaching an affine function of the same form. The coefficients themselves are small, and can be stored directly. Determining whether the subtraction result is negative or not is a problem of the same kind as was solved earlier: subtraction, multiplication and addition need to be calculated over variables; in this case the variables are the coefficients, instead of the original formal variables.

However, there is a distinct difference in working with coefficients, in that they, themselves, are calculable as polynomials over formal variables. The calculation can, therefore, be transformed into addition, multiplication and subtraction, once again over the original formal variables.

Although it may seem as though this conclusion returns us to the original problem, it does not. Consider, among all formal variables, the one defined last. This variable cannot appear in the exponentiation coefficients of any of the new polynomials. Therefore, the new equation is of the same type as the old equation but with at least one formal parameter less. Repeating the process over at most $n$ recursion steps (a polynomial number) allows us to compare any two indices for their sizes. $\square$

See the function `cmp` and the method `Command::cmp` in Appendix A for an implementation.

*Proof of Theorem 1.* We begin by recalling that P-RAM$[+, \leftarrow, \textit{Bool}] = $ PSPACE was already shown in [12]. Hence, we only need to prove P-RAM$[+, \dot{-}, \times, \leftarrow, \rightarrow, \textit{Bool}] \subseteq $ PSPACE. This is done, as per Schönhage's method [11], by simulating a polynomial time SLP$[+, \dot{-}, \times, \leftarrow, \rightarrow, \textit{Bool}]$ on a polynomial space Turing machine.

Lemmas 1–4, jointly, demonstrate that this can be done. $\square$

We remark that Theorem 1 is a striking result, in that right shifting is part of the SLP being simulated, and right shifting is a special case of integer division. Compare this with the power of exact division, described in Theorem 3, which is also a special case of integer division.

## 2.3 Incorporating arbitrary numbers

The framework described in Section 2.1 can readily incorporate simulation of arbitrary large number computation. We use it now, to prove Theorem 4.

*Proof of Theorem 4.* Having proved Theorem 1, what remains to be shown is

$$\text{PTIME-ARAM}[+, \dot{-}, \times, \leftarrow, \rightarrow, \textit{Bool}] \subseteq \text{PSPACE}.$$

As in the proof of Theorem 1, it is enough to show that an SLP that is able to handle all operations can be simulated in PSPACE.

We begin by noting that because the PTIME-ARAM must work properly for all but a finite range of numbers as its ALN input, it is enough to show one infinite family of numbers that can be simulated properly. We choose $X = 2^\omega$, for any sufficiently large $\omega$. In the simulation, we treat this $X$ as a new formal variable, as was done with outputs of "$a \leftarrow b$" operations.

Lemmas 1–3 continue to hold in this new model. They rely on the ability to compare between two indices, which, in the previous model, was guaranteed by Lemma 4. The technique by which Lemma 4 was previously proved was to show that comparison of two indices is tantamount to evaluating the sign of an affine combination of the exponents associated with a list of formal variables, when using their standard evaluation. This was performed recursively. The recursion was guaranteed to terminate, because at each step the new affine combination must omit at least one formal variable, namely the last one to be defined. Ultimately, the sign to be evaluated is of a scalar, and this can be performed directly.

When adding the new formal variable $X = 2^\omega$, the same recursion continues to hold, but the terminating condition must be changed. Instead of evaluating the sign of a scalar, we must evaluate the sign of a formal expression of the form $a\omega + b$. For a sufficiently large $\omega$ (which we assume $\omega$ to be), the sign is the result of lexicographic evaluation. □

A full C++ implementation of the solution appears in Appendix B.

## 3 Models with division

Our proof of Theorem 3 appears in Appendix C. It resembles [13] in that it uses Simon's ingenious argument that, for any given $n$, the value $\sum_{i=0}^{2^n-1} i \times 2^{ni}$ can be calculated in O(1)-time by considering geometric series summation techniques. The result is an integer that includes, in windows of length $n$ bits, every possible bit-string of length $n$. The simulating RAM acts by verifying whether any of these bit-strings is a valid tableau for an accepting computation by the simulated TM. This verification is performed using bitwise Boolean operations, in parallel over all options. The most salient differences between the proofs, being the errors in Simon's original argument that this paper corrects, are as follows.

1. Simon does not show how a TM can simulate an arbitrary RAM in ER-time, making his result a lower-bound only.

2. Simon uses what he calls "oblivious Turing machines" (which are different than those of [7]) in a way that simultaneously limits the TM's tape size and maximum execution time (only the latter condition being considered in the proof), and, moreover, are defined in a way that is non-uniform, in the sense that adding more tape may require a different TM, with potentially more states, a fact not accounted for in the proof.

3. Most importantly, Simon underestimates the length needed for the tableau, taking it to be the value of the input. TMs are notorious for using up far more tape than the value of their inputs (see [9]).

Ultimately, Theorem 3 proves that the power of a RAM[$op$], where $op$ is RAM-constructable and includes $\{+, /, \leftarrow, Bool\}$, is limited only by the maximal size of values that it can produce (relating to the maximal tableau size that it can generate and check). Considering this, the proof of Theorem 5 becomes a trivial corollary. The full details are given in Appendix D, but the basic idea is that any accepting computation by a TM is necessarily of some finite length $t$. Consider an operation set $op = \{+, /, \leftarrow, Bool, A\}$, where $A()$ is a function that has no parameters and returns a number that is at least as large as $t$. This places the computation in $\mathrm{EL}_{op}(\mathrm{O}(1))$-TM, so by Theorem 3 it is in O(1)-RAM[$op$], which is a subset of O(1)-ARAM[$+, /, \leftarrow, Bool$].

We have shown, therefore, that while arbitrary numbers have no effect on computational power without division, with division they provide Turing completeness in O(1) computational resources.

# A  Code to simulate P-RAM in PSPACE

## A.1  slp.h

```
/*
Interface file to the SLP and related classes.

This class simulates a PTIME-SLP[+,-,*,<<,>>,Bool] in PSPACE on a Turing machine.
*/

#include <algorithm>
#include <vector>

class SLP;                                                                    10
class Command;

class Index
{
 public:
   Command* command;
   std::vector<Command*> values;
   std::vector<int> counters;
   std::vector<int> lines;
   std::vector<int> maxima;                                                   20
 public:
   Index begin() const;
   Index end() const;
   Index rbegin() const;
   Index zero() const;
   Index unity() const;
   Index& next();
 public:
   Index() {}
   Index(Command* _command, const Index& old);                                30
   void make_index(int line, Command& value);
   Index& operator++();
   Index& operator--();
   Index normalize(const Index& ref) const;
};
```

```cpp
int cmp(const Index& index1, const Index& index2);

bool operator<(const Index& index1, const Index& index2);
bool operator>(const Index& index1, const Index& index2);
bool operator==(const Index& index1, const Index& index2);
bool operator>=(const Index& index1, const Index& index2);
bool operator<=(const Index& index1, const Index& index2);
bool operator!=(const Index& index1, const Index& index2);

Index operator+(const Index& index1, const Index& index2);
Index operator-(const Index& index1, const Index& index2);

class Command
{
  private:
    int m_line;
    Index m_begin;
    Index m_end;
    Index m_rbegin;
  protected:
    virtual int eval(const Index& index) const=0;
    Index make_index(Command& exponent);
    SLP* slp;
  public:
    const Index& begin() const { return m_begin; }
    const Index& end() const { return m_end; }
    const Index& rbegin() const { return m_rbegin; }
    virtual int cmp(const Index& index1, const Index& index2) const;
  public:
    Command(SLP& _slp);
    void set_slp(SLP* _slp, int _line);
    int operator[](const Index& index);
    int is_nonzero();
    int line() { return m_line; }
};

class Const : public Command
{
  private:
    int val;
  protected:
    virtual int eval(const Index& index) const;
  public:
    virtual int cmp(const Index& index1, const Index& index2) const;
    Const(SLP& _slp, int _val) : Command(_slp), val(_val!=0) {}
};

class And : public Command
{
  private:
    Command& arg1;
    Command& arg2;
```

```cpp
  protected:
    virtual int eval(const Index& index) const;                                90
  public:
    And(SLP& _slp, int _arg1, int _arg2);
};

class Or : public Command
{
  private:
    Command& arg1;
    Command& arg2;
  protected:                                                                   100
    virtual int eval(const Index& index) const;
  public:
    Or(SLP& _slp, int _arg1, int _arg2);
};

class Not : public Command
{
  private:
    Command& arg;
  protected:                                                                   110
    virtual int eval(const Index& index) const;
  public:
    Not(SLP& _slp, int _arg);
};

class Add : public Command
{
  private:
    Command& arg1;
    Command& arg2;                                                             120
  protected:
    virtual int eval(const Index& index) const;
  public:
    Add(SLP& _slp, int _arg1, int _arg2);
};

class Sub : public Command
{
  private:
    Command& arg1;                                                             130
    Command& arg2;
  protected:
    virtual int eval(const Index& index) const;
  public:
    Sub(SLP& _slp, int _arg1, int _arg2);
};

class Mult : public Command
{
  private:                                                                     140
    Command& arg1;
```

```cpp
    Command& arg2;
    int make_pn(Command& command, const Index& index) const;
  protected:
    virtual int eval(const Index& index) const;
  public:
    Mult(SLP& _slp, int _arg1, int _arg2);
};

class LShift : public Command                                                    150
{
  private:
    Command& mantissa;
    Index exponent;
  protected:
    virtual int eval(const Index& index) const;
  public:
    LShift(SLP& _slp, int _mantissa, int _exponent);
};
                                                                                 160
class RShift : public Command
{
  private:
    Command& mantissa;
    Index exponent;
  protected:
    virtual int eval(const Index& index) const;
  public:
    RShift(SLP& _slp, int _mantissa, int _exponent);
};                                                                               170

class SLP
{
  private:
    std::vector<Command*> commands;
  public:
    SLP();
    void push_back(Command* command);
    Command& operator[](int line) { return *commands[line]; }
    int is_nonzero() { return (*commands.rbegin())->is_nonzero(); }              180
    ~SLP();
};
```

## A.2  slp.cpp

/*
Implementation file of the SLP and related classes.

This class simulates a PTIME-SLP[+,-,*,<<,>>,Bool] in PSPACE on a Turing machine.
*/

```
#include "slp.h"
#include <numeric>

using namespace std;

Index Index::begin() const
{
  Index rc=*this;
  for(int i=0;i<maxima.size();++i) {
    rc.counters[i]=−maxima[i];
  }
  return rc;
}

Index Index::end() const
{
  Index rc=*this;
  rc.counters.clear();
  rc.values.clear();
  rc.maxima.clear();
  return rc;
}

Index Index::rbegin() const
{
  Index rc=*this;
  for(int i=0;i<maxima.size();++i) {
    rc.counters[i]=maxima[i];
  }
  return rc;
}

Index Index::zero() const
{
  Index rc=*this;
  for(int i=0;i<maxima.size();++i) {
    rc.counters[i]=0;
  }
  return rc;
}

Index Index::unity() const
{
  Index rc=zero();
  rc.counters[counters.size()−1]=1;
  return rc;
}

Index& Index::next()
{
  int i;
  for (i=0;i<maxima.size();++i) {
    ++counters[i];
```

```cpp
    if (counters[i]==maxima[i]+1) {
      counters[i]=-maxima[i];
    } else {
      return *this;
    }
  }
  if (i==maxima.size()) {
    *this=end();
  }
  return *this;
}

Index::Index(Command* _command, const Index& old)
  : command(_command), values(old.values), counters(old.counters),
    lines(old.lines), maxima(values.size())
{
  for(int i=0;i<values.size();++i) {
    maxima[i]=1<<(command->line()-lines[i]);
  }
}

void Index::make_index(int line, Command& value)
{
  values.push_back(&value);
  counters.push_back(0);
  lines.push_back(line);
  maxima.push_back(1<<(command->line()-line));
}

Index& Index::operator++()
{
  if (counters.size()==0) {
    return *this;
  }
  Index rc=end();
  for(Index i=begin();i!=end();i.next()) {
    if ((i>*this)&&((rc==end())||(rc>i))) {
      rc=i;
    }
  }
  *this=rc;
  return *this;
}

Index& Index::operator--()
{
  if (counters.size()==0) {
    return *this;
  }
  Index rc=end();
  for(Index i=begin();i!=end();i.next()) {
    if ((i<*this)&&((rc==end())||(rc<i))) {
      rc=i;
```

```cpp
    }
  }
  *this=rc;
  return *this;
}

Index Index::normalize(const Index& ref) const
{
  if (counters.size()==0) {
    return *this;
  }
  Index rc=ref.end();
  for(Index i=ref.begin();i!=ref.end();i.next()) {
    if ((*this>=i)&&((rc==ref.end())||(rc<i))) {
      rc=i;
    }
  }
  return rc;
}

int cmp(const Index& index1, const Index& index2)
{
  if (index1.counters.size()==0) {
    if (index2.counters.size()==0) {
      return 0;
    } else {
      return 2;
    }
  } else if (index2.counters.size()==0) {
    return −2;
  }
  Index indexL,indexR;
  if (index1.counters.size()>=index2.counters.size()) {
    indexL=indexR=index1;
  } else {
    indexL=indexR=index2;
  }
  for(int i=0;i<indexL.counters.size();++i) {
    indexL.counters[i]=indexR.counters[i]=0;
  }
  for(int i=0;i<index1.counters.size();++i) {
    indexL.counters[i]=max(index1.counters[i],0);
    indexR.counters[i]=max(−index1.counters[i],0);
  }
  for(int i=0;i<index2.counters.size();++i) {
    indexL.counters[i]+=max(−index2.counters[i],0);
    indexR.counters[i]+=max(index2.counters[i],0);
  }
  // comparing indexL to indexR is equivalent to comparing index1 to index2,
  // but none of the coefficients are negative.
  return indexL.command−>cmp(indexL,indexR);
}
```

```cpp
Index Command::make_index(Command& exponent)
{
  m_begin.make_index(line(),exponent);
  m_begin=m_begin.zero();
  m_end=m_begin.end();                                                                     170
  m_rbegin=m_begin.rbegin();
  return m_begin.unity();
}

int Command::cmp(const Index& index1, const Index& index2) const
{
  Command* command=(*index1.values.rbegin());
  Index i=command->begin();
  int maximum=max(accumulate(index1.counters.begin(),index1.counters.end(),0),
                  accumulate(index2.counters.begin(),index2.counters.end(),0));            180
  int logmax=0;
  while (1<<logmax<maximum) ++logmax;
  i.maxima[0]+=logmax;
  int acc1=0;
  int acc2=0;
  int sign=0;
  for(;i!=command->end();++i) {
    acc1>>=1;
    acc2>>=1;
    for(int j=0;j<index1.counters.size();++j) {                                            190
      int bit=(*index1.values[j])[i];
      acc1+=index1.counters[j]*bit;
      acc2+=index2.counters[j]*bit;
    }
    if (acc1%2>acc2%2) {
      sign=1;
    }
    if (acc1%2<acc2%2) {
      sign=-1;
    }                                                                                      200
  }
  return sign;
}

Command::Command(SLP& _slp)
{
  _slp.push_back(this);
}

void Command::set_slp(SLP* _slp, int _line)                                                210
{
  slp=_slp;
  m_line=_line;
  if (m_line>0) {
    m_begin=Index(this,(*slp)[m_line-1].begin());
    m_begin=m_begin.zero();
    m_end=m_begin.end();
    m_rbegin=m_begin.rbegin();
```

```
    }
    if (m_line==1) {                                                              220
      make_index(*this);
    }
}

bool operator<(const Index& index1, const Index& index2)
{
  return cmp(index1,index2)<0;
}

bool operator>(const Index& index1, const Index& index2)                          230
{
  return cmp(index1,index2)>0;
}

bool operator==(const Index& index1, const Index& index2)
{
  return cmp(index1,index2)==0;
}

bool operator>=(const Index& index1, const Index& index2)                         240
{
  return cmp(index1,index2)>=0;
}

bool operator<=(const Index& index1, const Index& index2)
{
  return cmp(index1,index2)<=0;
}

bool operator!=(const Index& index1, const Index& index2)                         250
{
  return cmp(index1,index2)!=0;
}

Index operator+(const Index& index1, const Index& index2)
{
  Index rc=index1;
  for(int i=0;i<rc.counters.size();++i) {
    rc.counters[i]=index1.counters[i]+index2.counters[i];
  }                                                                               260
  rc.normalize(index1);
  return rc;
}

Index operator-(const Index& index1, const Index& index2)
{
  Index rc=index1;
  for(int i=0;i<rc.counters.size();++i) {
    rc.counters[i]=index1.counters[i]-index2.counters[i];
  }                                                                               270
  rc.normalize(index1);
```

```
    return rc;
}

int Command::operator[](const Index& index)
{
  if (index<begin()) {
    return 0;
  }
  Index normalized=index.normalize(begin());                                         280
  return eval(normalized);
}

int Command::is_nonzero()
{
  for(Index index=begin();index!=end();++index) {
    if (eval(index)==1) {
      return 1;
    }
  }                                                                                  290
  return 0;
}

int Const::eval(const Index& index) const
{
  return val&&(index==begin());
}

int Const::cmp(const Index& index1, const Index& index2) const
{                                                                                    300
  return index1.counters[0]−index2.counters[0];
}

int And::eval(const Index& index) const
{
  return arg1[index]&&arg2[index];
}

And::And(SLP& _slp, int _arg1, int _arg2)
  : Command(_slp), arg1(_slp[_arg1]), arg2(_slp[_arg2])                              310
{
}

int Or::eval(const Index& index) const
{
  return arg1[index]||arg2[index];
}

Or::Or(SLP& _slp, int _arg1, int _arg2)
  : Command(_slp), arg1(_slp[_arg1]), arg2(_slp[_arg2])                              320
{
}

int Not::eval(const Index& index) const
```

```cpp
{
  int rc=1;
  for(Index i=rbegin();(i>=index)&&(rc==1);--i) {
    rc=1-arg[i];
  }
  if (rc==1) {
    return 0; // The tweaked NOT operator
  }
  return 1-arg[index];
}

Not::Not(SLP& _slp, int _arg) : Command(_slp), arg(_slp[_arg])
{
}

int Add::eval(const Index& index) const
{
  int acc=0;
  for(Index i=begin();i<=index;++i) {
    acc>>=1;
    acc+=arg1[i]+arg2[i];
  }
  return acc%2;
}

Add::Add(SLP& _slp, int _arg1, int _arg2)
  : Command(_slp), arg1(_slp[_arg1]), arg2(_slp[_arg2])
{
}

int Sub::eval(const Index& index) const
{
  int acc=0;
  Index i;
  for(i=begin();i<=index;++i) {
    acc>>=1;
    acc+=arg1[i]-arg2[i];
  }
  int rc=(acc%2)!=0;
  for(;i!=end();++i) {
    acc>>=1;
    acc+=arg1[i]-arg2[i];
  }
  if (acc==0) {
    return rc;
  } else {
    return 0; // arg2>arg1
  }
}

Sub::Sub(SLP& _slp, int _arg1, int _arg2)
  : Command(_slp), arg1(_slp[_arg1]), arg2(_slp[_arg2])
{
```

```
}

int Mult::make_pn(Command& command, const Index& index) const                          380
{
  int curr_bit=command[index];
  if (index<begin()) {
    return 0;
  } else if (index==begin()) {
    return −curr_bit;
  } else {
    Index prev=index;
    −−prev;
    int prev_bit=command[prev];                                                        390
    return prev_bit−curr_bit;
  }
}

int Mult::eval(const Index& index) const
{
  int acc=0;
  for(Index i=begin();i<=index;++i) {
    acc>>=1;
    for(Index p1=begin();p1!=end();++p1) {                                             400
      Index p2=i−p1;
      acc+=make_pn(arg1,p1)*make_pn(arg2,p2);
    }
  }
  return (acc%2)!=0;
}

Mult::Mult(SLP& _slp, int _arg1, int _arg2)
  : Command(_slp), arg1(_slp[_arg1]), arg2(_slp[_arg2])
{                                                                                      410
}

int LShift::eval(const Index& index) const
{
  return mantissa[index−exponent];
}

LShift::LShift(SLP& _slp, int _mantissa, int _exponent)
  : Command(_slp),
    mantissa(_slp[_mantissa]), exponent(make_index(_slp[_exponent]))                   420
{
}

int RShift::eval(const Index& index) const
{
  return mantissa[index+exponent];
}

RShift::RShift(SLP& _slp, int _mantissa, int _exponent)
  : Command(_slp),                                                                     430
```

```
    mantissa(_slp[_mantissa]), exponent(make_index(_slp[_exponent]))
{
}

SLP::SLP()
{
  new Const(*this,0);
  new Const(*this,1);
}

void SLP::push_back(Command* command)
{
  commands.push_back(command);
  (*commands.rbegin())->set_slp(this,commands.size()-1);
}

SLP::~SLP()
{
  for(vector<Command*>::iterator it=commands.begin();
      it!=commands.end();++it) {
    delete *it;
  }
}
```

## A.3 main.cpp

```
/*
Example program, using the class SLP.
*/

#include <iostream>

#include "slp.h"

using namespace std;

int main(void)
{
  SLP sample_slp;     // The SLP being simulated is: s[0] := 0; s[1] := 1;
  new LShift(sample_slp,1,1);        // s[2] := s[1] << s[1];
  new Sub(sample_slp,2,1);           // s[3] := s[2] - s[1];
  new Sub(sample_slp,3,1);           // s[4] := s[3] - s[1];
  cout << sample_slp.is_nonzero() << endl; // (s[4] != 0) ?
  return 0;
}
```

# B Code to simulate PTIME-ARAM in PSPACE

## B.1 aln.h

```
/*
Interface file to the ALN class.

This class extends the capabilities of class SLP by adding support of
arbitrary large numbers.
*/

#include "slp.h"

class ALN : public Command
{
  private:
    Index exponent;
  protected:
    virtual int eval(const Index& index) const;
  public:
    virtual int cmp(const Index& index1, const Index& index2) const;
    ALN(SLP& _slp) : Command(_slp), exponent(make_index(*this)) {}
};
```

## B.2 aln.cpp

```
/*
Implementation file of the ALN class.

This class extends the capabilities of class SLP by adding support of
arbitrary large numbers.
*/

#include "ALN.h"

int ALN::eval(const Index& index) const
{
  return index==exponent;
}

int ALN::cmp(const Index& index1, const Index& index2) const
{
  if (*index1.counters.rbegin()==*index2.counters.rbegin()) {
    Index indexL=index1;
    Index indexR=index2;
    indexL.counters.pop_back();
    indexL.values.pop_back();
    indexL.maxima.pop_back();
    indexR.counters.pop_back();
```

```
    indexR.values.pop_back();
    indexR.maxima.pop_back();
    return Command::cmp(indexL,indexR);
  } else {
    return (*index1.counters.rbegin())−(*index2.counters.rbegin());
  }
}
```
30

## B.3  main_w_ALN.cpp

```
/*
Example program, using the class ALN.
*/

#include <iostream>

#include "aln.h"

using namespace std;

int main(void)
{
  SLP sample_slp;      // The SLP being simulated is: s[0] := 0; s[1] := 1;
  new ALN(sample_slp);                 // s[2] := 2^\omega;
  new Add(sample_slp,1,2);             // s[3] := s[1] + s[2];
  cout << sample_slp.is_nonzero() << endl; // (s[3] != 0) ?
  return 0;
}
```
10

# C  Proof for Theorem 3

We prove Theorem 3 by first establishing a sequence of lemmas.

**Lemma 5.** *A TM can be simulated by a $RAM[\leftarrow, \rightarrow, Bool]$ using only bounded shifts.[2] A TM run requiring n execution steps can be simulated in this way by $O(n)$ RAM steps. In the simulation, advancing the TM by a single step is simulated by an $SLP[\leftarrow, \rightarrow, Bool]$.*

*Proof.* The simulation will store the state of the TM on three registers: *tape*, *head* and *state*. Register *tape* will store the current state of the tape. At start-up, this register is initialized by

$$tape \Leftarrow inp,$$

---
[2]That is, the right operand to the shift operation, being the exponent, is bounded by a value independent of the input. Equivalently, shifts can be restricted to shift-by-1. This is considered to be a weaker operation than general shifting.

where "$\Leftarrow$" is the assignment operator (as opposed to "$\leftarrow$", which signifies left shift).

Register *head* will store the position of the reading head. The format for doing so is that if the reading head is at position $i$ on the tape, the value of *head* will be $2^i$, this being a "1" at binary digit position $i$ and "0" everywhere else. At start-up, the register is initialized by

$$head \Leftarrow 1.$$

This places the head at the first position on the tape. (The count of position numbers begins at 0.)

Register *state* will signify the instantaneous state of the finite control. We number the states of the TM arbitrarily from 0 to $k-1$, where $k$ is the total number of states. By convention, state 0 will be the initial state of the machine. The format for storing the state number on the *state* variable is that the variable equals the state number times *head*. In other words, the number is stored starting at the bit position of the reading head. At start-up,

$$state \Leftarrow 0.$$

Let $c = \lceil \log_2 k \rceil$, this constant being the number of bits required to store the state number. The transition function of the Turing machine is a function from the current state and the value of the tape currently under the reading head to a new state, a new value and a new position of the reading head (which is at most one position from its previous position).

Consider the function only for the new value under the reading head. This is a function from $c+1$ bits to 1 bit, and can therefore be described by a finite number of Boolean operations. Applying this Boolean function requires, however, that all operands be available as bits. Consider, now, the numbers $state \rightarrow 0, \ldots, state \rightarrow (c-1)$. In each of these numbers a different bit of the original state description is aligned with the bit position of the reading head. Boolean algebra on these numbers along with the number *tape* will result in the output value having the correct new bit value for *tape* in the bit position that is under the reading head (though its other values may not signify anything meaningful). Let us refer to this numerical result as *output*.

To calculate the new value for *tape*, we note that the value under the reading head is the only one that requires updating. All other bit positions retain their original values. Hence,

$$tape \Leftarrow (output \wedge head) \vee (tape \wedge \neg head)$$

is the correct update.

We remark that the above expression relates to standard Boolean algebra. In our case, the operation "$\neg$" has been tweaked so as to ensure that the result is a nonnegative integer. (Non-tweaked negation would result in a Boolean string that has an infinite number of leading "1" bits, which would not be a valid result.) That being the case, $X \wedge \neg Y$ should be taken as a single (not tweaked) Boolean operation. To avoid confusion we denote it "$X$ clear $Y$" throughout, and continue to use "$\neg$" to mean tweaked negation.

Computing the value of the $i$'th bit of the new state number is attained by the same means as computing the new tape bit value. Let $outstate_i$ denote the result of applying Boolean operations on *tape* and $state \rightarrow 0$ through $state \rightarrow (c-1)$ to calculate this $i$'th bit. The update of *state* is

$$state \Leftarrow ((outstate_0 \wedge head) \leftarrow 0) \vee \cdots \vee ((outstate_{c-1} \wedge head) \leftarrow (c-1)).$$

Lastly, we need to update the variable *head*. To do so, we calculate three functions from *tape* and $state \to 0$ through $state \to (c-1)$, namely *IsRightMotion*, *IsStay* and *IsLeftMotion*, calculating whether the head is to move to the right, stay in place, or move to the left, respectively. The update required is

$$head \Leftarrow ((\textit{IsLeftMotion} \land \textit{head}) \leftarrow 1) \lor (\textit{IsStay} \land \textit{head}) \lor ((\textit{IsRightMotion} \land \textit{head}) \to 1).$$

This simulation, as described above, is already correct. However, we will change it slightly in order to make it more elegant. Specifically, we wish to avoid the scenario of the reading head dropping off the edge of the tape, which may happen if $head = 1$ and the next motion is a motion to the right. The result will be that the new *head* value is $0$.[3]

To avoid this, we introduce a new element into the simulation. This is a constant. Its name is *boundary* and its value is 1. Furthermore, we add one new state into the state machine, this being a rejecting halting state, signifying that the tape head has dropped off the tape. (Adding a new state may cause an increase in $c$, requiring a change in the update functions.) The boundary condition is a new bit in the input of the update functions. Its value is $boundary \land head \land \textit{IsRightMotion}$, and it triggers a transition to the new rejecting state, and ultimately no head motion. □

**Lemma 6.** *A TM working on a bounded tape of length $s$ can be simulated by a $RAM[\leftarrow, \to, Bool]$ using only bounded shifts if the RAM is given $B = 1 \leftarrow s$ as an additional input. The simulation is required to be uniform in this input. A TM run requiring $n$ execution steps can be simulated in this way by $O(n)$ RAM steps. In the simulation, advancing the TM by a single step is simulated by an $SLP[\leftarrow, \to, Bool]$.*

Note that the parameter $s$ is not given as an input to the RAM.

*Proof.* We want to make the simulation of Lemma 5 into a tape-bounded one. In the new simulation, the value of *boundary* will be $1 \lor (B \leftarrow (c-1))$.

To the boundary condition discussed in Lemma 5 we now add a second, triggered by $head \land \textit{IsLeftMotion} \land (boundary \to c)$, which will cause the state machine to transition to another new halting state, namely one signifying that the simulation was stopped due to the reading head exceeding its tape allocation. □

An explanation is due as to the question of why the second 1 bit of the boundary is at position $s+c-1$, and not at position $s$. The reason is that *boundary* does not signify the end of the tape, but rather the end of all bits used in the simulation process. When the reading head is at its left-most position (position $s-1$) the state of the TM is written in parameter *state* between its $s-1$'th and $s+c-2$'th position. Therefore, the $s+c-1$ bit is the first not to be part of the simulation. Some of the bits before the $s+c-1$ bit, such as the last $c-1$ bits of *tape* are not used in the simulation process.

This property is put to use in Lemma 7.

---

[3]The TM being simulated may legitimately move the reading head beyond the edge of the tape as a method of rejecting the input.

**Lemma 7.** *For any nonnegative $k$, and any naturals $s_1, \ldots, s_k$, $k+1$ copies of the same TM, working on inputs $inp_1, \ldots, inp_{k+1}$ on tapes of sizes $s_1, \ldots, s_k$ and infinity, respectively, can be simulated by a $RAM[\leftarrow, \rightarrow, Bool]$ using only bounded shifts, given that the RAM is given its input in the following format:*

$$B = \sum_{i=0}^{k} 2^{\sum_{j=1}^{i}(s_j+c-1)},$$

$$inp = \sum_{i=0}^{k} \left( inp_{i+1} \leftarrow \left( \sum_{j=1}^{i}(s_j+c-1) \right) \right),$$

*and that the output is described by*

$$output = \sum_{i=0}^{k} \left( output_{i+1} \leftarrow \left( \sum_{j=1}^{i}(s_j+c-1) \right) \right),$$

*where $output_1, \ldots, output_{k+1}$ are the halting states of the $k+1$ TMs at program termination. The simulation requires $O(n)$ RAM steps, where $n$ is the number of steps required by the longest running of the executions. In the simulation, advancing all the TMs together by a single step is simulated by an $SLP[\leftarrow, \rightarrow, Bool]$.*

*Proof.* We simulate all TMs in parallel. In doing so, it is convenient to simulate an "advance one step" also on the TMs that have already halted. Let us therefore reconsider the term "halting state", in order for us to allow continued execution of the simulation even after a halt. To allow this, transition functions need to be defined also for halting states. We define these as follows: if $x$ is a halting state, transitions from $x$ are all to $x$, the tape contents are never changed, and the reading head will always move to the right, unless it is already at the end of the tape, in which case it will stop. Clearly, all this can be encapsulated by the Boolean functions already discussed.

We note that if a TM runs for $n$ steps, its reading head is no more than $n$ positions from the right end of the tape. Hence, in $O(n)$ steps, the TM's state, head-position and tape contents will have reached their final values.

The reason for the strange behavior we require for the halting states is that it allows us to easily query the TM simulator, at the end of the simulation, regarding its halting condition. Let us assume, without loss of generality, that the only halting states are

- ***state* = 1** signifying an accepting calculation,

- ***state* = 2** signifying a rejecting calculation, and

- ***state* = 3** signifying that the calculation was aborted due to exceeded tape length.

Because the ultimate position of the reading head is known to be 0, querying the simulator for the final state is simply a test for equality.

To initialize the simulator we set

$$boundary \Leftarrow B,$$
$$head \Leftarrow B,$$
$$state \Leftarrow 0,$$
$$tape \Leftarrow inp.$$

Because the positions of the "1"s on *boundary* signify a separation of the bit positions into segments that cannot interact, simply using the SLP$[\leftarrow, \rightarrow, Bool]$ described in Lemma 5 and then re-used in Lemma 6 results in all TMs advancing one step in parallel.

Lastly, we need to test whether all TMs have halted. This can be split into the following conditions. First, we verify that
$$head = B$$
and
$$state \text{ clear } (B \vee (B \leftarrow 1)) = 0.$$

This guarantees that all TMs are either in one of the three halting states or in the initial state, 0. To verify that none are in zero, we simply check that the set of TMs that have halted in one of the halting conditions covers all TMs. This is done by

$$(state \vee (state \rightarrow 1)) \wedge B = B.$$

When all of the above conditions are satisfied, execution of the RAM terminates, and the output can be read off *state*. □

We remark regarding Lemma 7 that even though parallelization over $k+1$ machines is possible, it is not necessary. By zeroing-out all bits of *head* in any given segment, the TM related to that segment ceases to advance. Indeed, that would have been a different method to approach the question of how to simulate the execution of machines that have already halted.

We now strengthen Lemmas 5, 6 and 7 further, by omitting from each the right-shift operator, which can be done by means of the following lemma.

**Lemma 8.** *For op = $\{\leftarrow, \rightarrow, [+], [\dot{-}], Bool\}$, if a RAM[op] does not use indirect addressing and is restricted to bounded shifts, it can be simulated by a RAM[op\\{$\rightarrow$}] without loss in time complexity. This result remains true also if the RAM can apply "$a \rightarrow b$" when b is the (unbounded) contents of a register, provided that the calculation of b does not involve use of the "$\rightarrow$" operator.*

*Proof.* We begin by considering the case of bounded shifts.

A RAM that does not use indirect addressing is inherently able to access only a finite set of registers. Without loss of generality, let us assume that these are $R[0], \ldots, R[k]$. The simulating RAM will have $R'[0], \ldots, R'[k+1]$ satisfying the invariant

$$\forall i : 0 \leq i \leq k, R[i] = R'[i]/R'[k+1].$$

To do this, we initialize $R'[k+1]$ to be 1, and proceed with the simulation by translating any action by the simulated RAM on $R[i]$, for any $i$, to the same action on $R'[i]$.[4] We do this for all actions except $R[i] \to X$, which is an operation that is unavailable to the simulating RAM.

To simulate "$R[j] \Leftarrow R[i] \to X$", we perform the following.

1. $R'[j] \Leftarrow R'[i]$.

2. $\forall x : x \neq j, R'[x] \Leftarrow R'[x] \leftarrow X$.

3. $R'[j] \Leftarrow R'[j]$ clear $\neg R'[k+1]$.

We note regarding the second step that this operation is performed also on $x = k+1$. The fact that $k$ is bounded ensures that this step is performed in O(1) time.

Essentially, if "$R[j] \Leftarrow R[i] \to X$" is thought of as "$R[j] \Leftarrow \lfloor R[i]/2^X \rfloor$", Step 1 performs the assignment, Step 2 the division, and Step 3 the truncation.

In order to support "$R[j] \Leftarrow R[i] \to X$" also when $X$ is the product of a calculation, the simulating RAM also performs, in parallel to all of the above, a direct simulation that keeps track of the register's native values. In this alternate simulation, right shifts are merely ignored. Any calculation performed by the simulated RAM that does not involve right shifts will, however, be calculated correctly, so the value of $X$ will always be correct. □

Before proceeding further, we introduce three definitions.

**Definition 4** (Instantaneous Description). Let $\mathcal{T}$ be a TM working on a bounded tape of size $s$ and having a state space that can be described (including up to 3 additional halting states, as in Lemma 7) by $c$ bits.

An *instantaneous description* of $\mathcal{T}$ at any point in its execution is the value of $tape + 2^{s+c-1} state + 2^{2(s+c-1)} head$ at that point in its execution, where *tape*, *state* and *head* are as in Lemma 6.

**Definition 5** (Vectors). A triplet $(m, V, n)$ of integers will be called an *encoded vector*. We refer to $m$ as the *width* of the vector, $V$ as the *contents* of the vector and $n$ as the *length* of the vector. If $V = \sum_{i=0}^{n-1} 2^{mi} k_i$ with $\forall i : 0 \leq i < n \Rightarrow 0 \leq k_i < 2^m$, then $[k_0, \ldots, k_{n-1}]$ will be called the *vector* (or, the *decoded vector*), and the $k_i$ will be termed the *vector elements*. Notably, vector elements belong to a finite set of size $2^m$ and are not general integers. It is well-defined to consider the most-significant bits (MSBs) of vector elements. Nevertheless, any $n$ integers can be encoded as a vector, by choosing a large enough $m$.

Actions described as operating on the vector are mathematical operations on the encoded vector (typically, on the vector contents, $V$). However, many times we will be more interested in analyzing these mathematical operations in terms of the effects they have on the vector elements. Where this is not ambiguous, we will name vectors by their contents. For example, we can talk about the "decoded $V$" to denote the decoded vector corresponding to some encoded vector whose contents are $V$.

---

[4] An action involving an explicit "1" (except for shifting by 1) will have the "1" replaced by $R'[k+1]$ in the simulation.

**Definition 6** (Tableau). Let $\mathcal{T}$, $s$ and $c$ be as in Definition 4. A *tableau* is a vector of width $3(s+c-1)$, whose $i$'th element is the instantaneous description of $\mathcal{T}$ after $i$ execution steps.

Furthermore, we define

$$O_{a,m}^n \stackrel{\text{def}}{=} ((a \leftarrow nm) \mathbin{\dot{-}} a)/((1 \leftarrow m) \mathbin{\dot{-}} 1).$$

For $a < 2^m$, this is the vector $[a, \ldots, a]$.

**Lemma 9.** *There exists a constant, $K$, such that for every TM, $\mathcal{T}$, there exists a RAM$[\leftarrow, \text{Bool}]$, $\mathcal{R}$, that takes $K+1$ inputs, such that for every number inp, $\mathcal{T}$ accepts inp if and only if there exist witness numbers $w_1, \ldots, w_K$, for which $\mathcal{R}$ accepts on*

$$(\mathit{inp}, w_1, \ldots, w_K),$$

*and $\mathcal{R}$ executes in $O(1)$ time on every set of inputs.*

*Proof.* The TM $\mathcal{T}$ accepts if and only if there is an accepting computation path beginning at the initial instantaneous description at the start of execution with *inp* as its input. (The terminology "accepting computation path" is more commonly applied to nondeterministic computation. In deterministic computation, there is exactly one path. The question is only whether it is accepting or not.) If there exists an accepting computation path, it can be described by a tableau. (If it does not exist, no corresponding tableau exists.) Let $(m, V, n)$ be this tableau. We will choose witness integers so that the RAM is able to verify the existence and correctness of the tableau.

We use $K = 5$. The auxiliary inputs to the RAM will be the following five witness integers.

$$\begin{aligned}
w_1 &= m/3, \\
w_2 &= V, \\
w_3 &= (n-1)m, \\
w_4 &= O_{2^{m/3}-1,m}^n, \\
w_5 &= O_{1,m}^n.
\end{aligned}$$

The triplet $(w_1, w_2, w_3)$ defines the tableau, by a reversible transformation. The first step of the verification process is not to test the validity of this tableau, but rather to ascertain that $w_4$ and $w_5$ correspond properly to this tableau definition.

Let $a = 2^{m/3} - 1$, and note that it can be computed by $\neg(1 \leftarrow w_1)$. Ascertaining the validity of $w_4$ is done by verifying that

$$w_4 - a = (w_4 - (a \leftarrow w_3)) \leftarrow w_1 \leftarrow w_1 \leftarrow w_1.$$

This follows directly from the definition of $O_{2^{m/3}-1,m}^n$.

We do not actually have subtraction as part of our operation set, but $A - B$ can be simulated in this case by $A$ clear $B$ after having verified that $A \wedge B = B$.

Validating $w_5$ is done following exactly the same principles, only replacing $a$ for 1 in the computation.

At first glance, it may seem that $w_3$ also needs to be verified in order to ascertain that it is a multiple of $3w_1$, but in fact this has already been covered by the previous checks: if $2^x - 1$ is a multiple of $2^y - 1$ then $x$ is a multiple of $y$. That being the case, the fact that we have already established
$$w_5 = \frac{2^{w_3+3w_1} - 1}{2^{3w_1} - 1}$$
directly implies that $w_3 + 3w_1$ is a multiple of $3w_1$, and therefore also $w_3$ is a multiple of $3w_1$.

With these verified, the vector contents of the tableau, $V$, given in $w_2$, can be decomposed into its constituents.
$$tape = (V \wedge w_4) \leftarrow w_1 \leftarrow w_1,$$
$$state = (V \wedge (w_4 \leftarrow w_1)) \leftarrow w_1,$$
$$head = V \wedge (w_4 \leftarrow w_1 \leftarrow w_1).$$

Furthermore, a corresponding *boundary* parameter can be devised by
$$boundary = (w_5 \leftarrow w_1 \leftarrow w_1) \vee (w_5 \leftarrow w_1 \leftarrow w_1 \leftarrow w_1).$$

On the assumption that the tableau is legitimate, if we were to use these new values as inputs for the RAM described in Lemma 7 (which can be used here, because the right-shift operations can be circumvented, as per Lemma 8), then what these inputs describe is $n$ copies of the same TM, each performing execution on the same input, but at a different stages in the execution. The first TM is at execution start, the next one is after one execution step, etc.. (Note: the tableau is shifted left by $2w_1$, but, as stated before, this merely introduces an "unused" machine into the proof of Lemma 7.)

We can now use the RAM described in Lemma 7 to simulate, in $O(1)$ total time, one execution step of each of these $n$ TM steps. Let us denote the output variables after a single-step operation ($tape'$, $state'$, $head'$). The legitimacy of the tableau can now be verified. A true tableau will satisfy two conditions:

1. The state of the machine described in element 0 of the original tableau is the correct initial state of the TM to be simulated.

2. For each $i$, the instantaneous description of the machine described in element $i$ of the tableau, after advancing by a single step, equals the original instantaneous description in element $i+1$ of the tableau.

Verifying these conditions for *tape*, for example, is done by checking that
$$(inp \wedge a) \vee ((tape' \text{ clear } (a \leftarrow w_3)) \leftarrow w_1 \leftarrow w_1 \leftarrow w_1) = tape.$$

The idea is that $tape' \leftarrow m$ should equal $tape$ in all but the first and last elements. These are handled separately. The same method works for verifying the validity of *state* and *head*.

Once the tableau is known to be correct, we finish by ascertaining that its final state is an accepting state:
$$head \wedge (a \leftarrow w_3 \leftarrow w_1 \leftarrow w_1) = 1 \leftarrow w_3 \leftarrow w_1 \leftarrow w_1$$

and
$$state \wedge (a \leftarrow w_3 \leftarrow w_1 \leftarrow w_1) = 1 \leftarrow w_3 \leftarrow w_1 \leftarrow w_1$$

□

Though not required for the main proofs that these lemmas have all been building up towards, it is still illuminating to see that Lemma 9 implies a much stronger claim regarding nondeterministic computation. This is discussed in Appendix E.

We now turn to the main proof.

*Proof of Theorem 3.* The statement of the theorem involves equality between four Turing machine related complexities and one RAM complexity. The equivalence between the deterministic time and deterministic space TM complexities stems from the well-known argument that

$$\text{TIME}(n) \subseteq \text{SPACE}(n) \subseteq \text{TIME}(\text{EXP}(n)).$$

In $n$ execution steps, the reading head cannot reach more than $n$ elements into the tape, proving $\text{TIME}(n) \subseteq \text{SPACE}(n)$, whereas an execution bounded by $n$ tape elements cannot proceed more than $\text{EXP}(n)$ execution steps without retracing its steps, proving $\text{SPACE}(n) \subseteq \text{TIME}(\text{EXP}(n))$. In expansion limit terms, $2^{\text{EL}_{op}(n)}$ is greater than $\text{EL}_{op}(n)$ but is bounded from above by $\text{EL}_{op}(n+\text{O}(1))$. This argument shows equivalence between $\text{EL}_{op}(n + \text{O}(1))$ time-bounded TMs and $\text{EL}_{op}(n + \text{O}(1))$ space-bounded TMs, which is a stricter condition than the one actually needed for the theorem.

The addition of nondeterminism to a space-bounded TM requires only a limited amount of extra space to simulate on a deterministic TM, as is proved by Savitch's Theorem [10], so here, too, an $\text{EL}_{op}(n + \text{O}(1))$ is all that is needed. Simulating an $n$ time nondeterministic TM by a deterministic TM requires at most $n$ nondeterministic bits, and can hence be simulated by $\text{EXP}(n)$ runs of the deterministic algorithm, enumerating over the witness string. Exponentiation is, again, within $\text{EL}_{op}(n + \text{O}(1))$.

We will, therefore, only need to prove the following equality.

$$\text{O}(f(n))\text{-RAM}[op] = \text{EL}_{op}(\text{O}(f(n)))\text{-SPACE-TM}.$$

This, in turn, can be thought of as two statements. We begin by demonstrating the simpler

$$\text{O}(f(n))\text{-RAM}[op] \subseteq \text{EL}_{op}(\text{O}(f(n)))\text{-SPACE-TM}, \tag{1}$$

then continue to our main argument,

$$\text{O}(f(n))\text{-RAM}[op] \supseteq \text{EL}_{op}(\text{O}(f(n)))\text{-SPACE-TM}. \tag{2}$$

To prove Equation (1) we show that an $\text{EL}_{op}(\text{O}(f(n))$ space-bounded TM can simulate an $\text{O}(f(n))$-RAM. We allow this simulation to use a larger set of tape symbols than the original $\{0, 1\}$, knowing that this costs at most a linear increase in the amount of space required [15], which cannot violate the $\text{EL}_{op}(\text{O}(f(n))$ bound. Furthermore, as per [3], we continue to consider only RAMs that make no use of indirect addressing (and are therefore known to only access a finite number of registers).

Consider the following memory layout which may be used by the TM. Each of the (finite) number of registers of the RAM is allocated a new tape-alphabet character. We refer to these as "control" characters. During the TM's execution, each control character will appear on the tape exactly once. All characters to the right of it and to the left of the next control character are deemed to be the contents of the relevant register. Additionally, a further control character, appearing last on the tape, marks the start of a "scratchpad". The TM is initialized by moving the input one element to the right in order to make room for the $R[0]$ control character, then writing the $R[0]$ control character before the input and the rest of the control characters after (signifying registers whose contents are zero).

At each execution step, the operands are copied to the scratchpad, the operation required by the RAM is performed, the tape elements are moved enough to allocate enough space on the target register, and then the scratchpad result is copied to the target register and the scratchpad is cleared.

By the definition of EL, each register's contents requires at most $\text{EL}(f(n))$ elements to store. To store all registers we require $O(\text{EL}(f(n)))$. By definition of RAM-constructability, the scratchpad requires no more than $\text{EL}(O(f(n)))$ elements of memory.[5] Together, the amount of tape required is bounded by $\text{EL}(O(f(n)))$, as required by the Theorem.

We now turn to the main problem of proving Equation (2). We do this by working from Lemma 9, noting that if there exists a tableau computation $(m, V, n)$, then there also exists one with $(m', V, n')$, where $m'$ can be any number at least as large as $m$, and $n'$ is determined from $m'$ so as to be at least as large as the number of possible instantaneous descriptions for the TM. (If the TM requires more steps, it will necessarily have retraced its steps and have entered an infinite loop.)

Given a chosen value for $m'$, the computed value of $n'$ allows us to determine the number of bits in $V$. It is $T = n'm'$. $V$ is, therefore, one of finitely many possibilities. If we were to enumerate over all $2^T$ of these possibilities, we would be able to answer whether a tableau for an accepting computation exists for a particular value of $m'$.

Consider now several tableau candidates, $tableau_0, \ldots, tableau_{k-1}$, each of which is a variable in the range $0 \leq v < 2^T$. We wish to check simultaneously which of these (if any) is a correct tableau of bit-length $T$, describing the execution of a TM running on a tape of length $s$ on input $inp$. To do this, we describe a new variable, $V = \sum_{i=0}^{k-1} 2^{Ti} tableau_i$.

Consider now Lemma 7. It allows the verification of a tableau by use of only bitwise operations and shifts by values that are functions only of $s$. Because all candidates share the same $s$, each of these operations is executed on all candidates simultaneously, by executing them on $V$. There are only two places where this strategy fails.

First, the verification requires the use of several constants, such as 1 and $inp$. In order to apply these simultaneously to all simulations, we need to replace the original constants by $O_{1,T}^k$ and $O_{inp,T}^k$, respectively. Second, the final step in verifying whether a tableau candidate is correct involves several comparisons of the type "$a = b$". These must now be executed in parallel, for which we need an operation EQ that takes two vectors and outputs a new vector of the same length and width as its two operands, whose elements are 1 in positions where $a_i = b_i$ and 0 otherwise.

---

[5]This follows from $M_{op}(a+b, inp) = \max_x M_{op}(b, x))$, where the maximum is taken over all $x$ that are computable from $inp$ in $a$ calculation steps.

Appendix F describes how both of these can be calculated in O(1) time given the available set of operations, assuming $k = 2^T$.

A correct tableau is one that passes all equality checks outlined in the proof of Lemma 7. To simulate this in parallel for all candidates, we perform the necessary EQ per equality test, then take the bitwise AND over all results. The resulting vector, *res*, has a 1 in its $Ti$'th position if and only if the $tableau_i$ candidate is correct.

For the purposes of the present proof, we are not interested in finding which of the tableau candidates is correct, but rather whether *any* of them are correct. This can be checked by a simple "$res = 0$".

(Readers may wish to verify that the proof of Lemma 7 does not depend on the tableau candidate verified starting at bit-position 0, nor is the verification process disrupted by having any additional non-zero bits to the left or right of the tableau bits. The verification process does not use or disturb any bits in the integer, other than the $T$ being checked.)

The above argument, showing that the condition "the tableau is correct" can be verified in constant time can be applied equally well with additional checks such as "the tableau is correct and the halting state is 1". Thus, if we were to check all $2^T$ tableau candidates, the program does not only work as a verifier, but also as a complete simulator: it takes an input, and can return (in constant time) what the final state of the TM finite control is. In our case, the interesting final states are 1, 2 and 3. The simulator should return which of these three the TM halted on, or return that none of the above occurred.

The only remaining question is how to create all $k = 2^T$ possible tableaus in a single integer, for them to be verified simultaneously. One way to do this is as follows. The function described produces the contents of a vector of width $T$ and length $2^T$, whose elements are all unique.

$$U^T = \sum_{i=1}^{2^T-1} O^i_{1,T} = \sum_{i=1}^{2^T-1} \frac{2^{Ti} - 1}{2^T - 1} = \frac{\frac{2^{T \times 2^T} - 1}{2^T - 1} - 2^T}{2^T - 1}$$
$$= ((\neg(1 \leftarrow (T \leftarrow T))/\neg(1 \leftarrow T)) \text{ clear } (1 \leftarrow T))/\neg(1 \leftarrow T).$$

It is computed solely by use of left-shifting, exact division and Boolean operations.

We see, therefore, that all tableaus can be created and verified for correctness in constant time. Let us refer to the RAM program that verifies simultaneously and in constant time all possible tableaus pertaining to the execution of a TM on a tape of size $s$ as $simulate(s, inp)$.

By the assumption of constructability, there also exists a RAM program, $el(n)$ that calculates a value no smaller than $EL_{op}(n)$ in O($n$) time. In order to complete the proof of the theorem, we therefore present Algorithm 1 that enables a RAM working in O($n$) time to simulate any TM working in EL(O($n$)) space.

We remark that Algorithm 1 accepts if the original TM accepts, and continues indefinitely if the original TM does not, but this is merely because we check only for a halting state of 1. It is also possible to terminate the run and reject if the halting state is 2 (signifying that the TM halted and rejected) or if the halting state is not in 1-3 (signifying that the TM returned to a previous instantaneous description, and has therefore entered an infinite loop). However, if the TM continues indefinitely while consuming unbounded amounts of tape, the simulation cannot detect this, and

**Algorithm 1** A complete TM simulator

1: $n \Leftarrow 1$
2: **loop**
3:    $s \Leftarrow el(n)$
4:    $simulate(s, inp)$
5:    **if** halting state is 1 **then return** Accept
6:    **end if**
7:    $n \Leftarrow n + n$
8: **end loop**

will continue forever. This is, of course, inherent, because had the simulator been able to determine in finite time for every TM whether it halts or not, this would have contradicted Turing's halting theorem [17]. □

## D   Adding arbitrary numbers

In the main paper we present the general reasoning that leads to Theorem 5. The formal argument is as follows.

*Proof of Theorem 5.* If $S$ is an r.e. set, then there is a TM that recognizes it, indicating that $inp \in S$ if and only if there is an accepting computation of the TM with $inp$ as the TM's input. This accepting computation by definition requires only a finite number of execution steps of the TM, and therefore only a finite number, $s_{req}$, of tape cells. Consider Algorithm 2, using the RAM-implementable function *simulate* discussed in Appendix C.

**Algorithm 2** Simulating a TM in constant time

1: $s \Leftarrow$ ALN
2: $simulate(s, inp)$
3: **if** halting state is 1 **then**
4:    **return** Accept
5: **else**
6:    **return** Reject
7: **end if**

By the arguments of the proof for Theorem 3, a sufficient condition for *simulate* to return the proper result is $s \geq s_{req}$. This is clearly satisfied by making $s$ "large enough", which is what is guaranteed by "$s \Leftarrow$ ALN". Formally, almost all choices for $s$ return the correct result, as the number of $s$ values that return an incorrect result is finite and bounded by $s_{req}$. □

# E  Results about nondeterministic computation

In Turing machines, nondeterministic computation is often described as computation that makes use of an extra tape, with Oracle-provided information. In Turing machine computation, a machine that runs for $n$ execution steps can necessarily access no more than $n$ bits of this extra tape, so the Oracle-provided information (sometimes referred to as a "witness" or a "certificate") for a polynomial-time TM algorithm is effectively limited to only a polynomial number of bits.

It has been shown (e.g. in [5, 8]) that the availability of such a certificate does not increase computational power in sufficiently-equipped RAMs. However, in [13] an alternative is discussed. RAMs have the ability to access entire integers in O(1)-time. Simon suggests that for RAMs one can analyze a variant of nondeterministic computation where the information provided by the Oracle is in the form of an integer of arbitrary length. We refer to a RAM with access to such an integer as an NRAM.

The following theorem, following from Lemma 9, strengthens Simon's results about NRAMs in [13].

**Theorem 6.** *All recursively enumerable (r.e.) sets can be recognized in $O(1)$ by an NRAM[op], where op is a superset of any of the following: $\{\leftarrow, inc, Bool\}$, $\{\rightarrow, inc, Bool\}$, $\{/, inc, Bool\}$ and $\{\times, inc, Bool\}$, where inc is the increment function.*[6]

We prove separately for each of the instruction sets in the theorem.

*Proof for $op \supseteq \{\leftarrow, inc, Bool\}$.* For this part of the proof, the missing piece from Lemma 9 is the ability to extract all 5 required integers, $w_1, \ldots, w_5$, from a single, Oracle-provided integer, $\alpha$, in O(1) time.

A critical observation is that there are some degrees of freedom in the choice of witnesses. For example, $w_1$ merely needs to be large enough. It signifies the length of the tape available to the TM, so choosing a larger value does not invalidate the simulation. As such, we can restrict $w_1$ to be a power of two at no cost.

Next, we note that $w_3$ is not actually needed from the Oracle. The number $w_3$ provides the length, $n$, of the tableau, but choosing an overly long tableau is not a problem. All that needs to be ascertained is that the chosen length is not too small.

Consider how many simulation steps are needed to decide how a TM's execution terminates. A TM on a bounded-length tape is essentially a finite state machine (FSM). It only has a constant number of instantaneous descriptions that it can be in. In our case, each instantaneous description is stored by $m$ bits. Hence, the size of the state space is trivially bounded by $S = 2^m$. There is no reason to simulate a deterministic FSM to more steps than the size of its state space, because if it has not terminated after $S$ steps this indicates that it has revisited a state more than once, and is therefore in an infinite loop.

Assuming that $w_1$ is a power of two, multiplication by 3 can be performed by

$$m \Leftarrow w_1 \vee (w_1 \leftarrow 1).$$

---

[6]This function is generally available to all RAMs and is often not listed on the operation set (c.f. [11]).

Using $w_3 \Leftarrow m \leftarrow m$ is then equivalent to choosing $S+1$ for $n$, which is more than enough.

This leaves us with the necessity to extract $w_1$, $w_2$, $w_4$ and $w_5$ from $\alpha$.

We do so by first considering inc($\alpha$) clear $\alpha$. If $\alpha$'s $u$ lowest bits are ones, but the $u+1$ bit is 0, this operation yields the number $2^u$. We will take this to be $w_1$, noting that by construction it is a power of two, as desired.

The next observation to make is that $w_2$, $w_4$ and $w_5$ all have at most $W = w_3 + m$ bits. As such, consider $\alpha$ as the contents of a vector of width $W$ and length 4. The vector element in the 0'th position, of which $u+1$ bits were already specified, we ignore. The remaining elements we take to be $w_2$, $w_4$ and $w_5$, respectively.

Let $M = 2^W$. This value can be calculated by

$$M \Leftarrow 1 \leftarrow w_3 \leftarrow w_1 \leftarrow w_1 \leftarrow w_1.$$

Given a right-shift operator, the three witness values would have been extractable by

$$w_2 = (\alpha \to w_3 \to w_1 \to w_1 \to w_1) \wedge \neg M,$$

and similarly for $w_4$ and $w_5$, from which point the rest of the construction could follow that of Lemma 9.

Because right-shifting is, in fact, not assumed to be part of *op*, we note, instead, that all shifts in both the construction above and that of Lemma 9 are by amounts that require no right-shifting to calculate. (Aside from shifts by a constant, they are by $w_1$, $w_3$ and $m$, which we show here how to calculate explicitly with no use of right-shifting.) That being the case, we can apply Lemma 8 to show that the right-shift operator is not necessary. □

*Proof for op $\supseteq \{\to, inc, Bool\}$.* When right-shifting is the only shift available, we employ a slightly different tactic. Given a right-shift operator, unpacking in O(1) time $K$ integers, $a_1, \ldots, a_K$ from a given integer, $\alpha$, is fairly straightforward. We do this by storing in $\alpha$ the contents of the vector $[u-1, a_1, \ldots, a_K]$, where $u$ is the width of the vector and is a power of two and greater than any of the $a_i$. (Regardless of what values we wish to store in the $a_i$, it is always possible to choose an appropriate width, $u$. We note that unlike in the first part of the proof, here $u$ does not serve any role other than being the width of this vector. It does not participate in the actual simulation of the TM.)

Finding $u$ is done by

$$u \Leftarrow \text{inc}(\alpha) \text{ clear } \alpha$$

as in the first part of the proof. Extracting the elements is done by

$$a_i = (\alpha \overbrace{\to u \cdots, \to u}^{i \text{ times}}) \wedge \neg u.$$

In order to proceed from $K$ extracted integers to a proof regarding a TM, we use the same algorithm as that of Lemma 9 (not using any of the short-cuts introduced for $op \supseteq \{\leftarrow, inc, Bool\}$). However, we switch every left shift for a right shift by the following technique.

First, note that Lemma 9 guarantees a procedure that terminates in O(1) time. As such, it can only use a bounded number of left-shift operations.

We can therefore transform each $X \Leftarrow Y \leftarrow Z$ operation in the original proof to a comparison step: $X = Y \leftarrow Z$. If all of $X, Y, Z$ and $W = 2^Z - 1$ are given as part of $\alpha$, this can be performed simply by the following verification steps.

First, we verify that $W + 1$ is a power of two by $\mathrm{inc}(W) \wedge W = 0$. Next, we verify that it is the correct power of two by $\mathrm{inc}(W) \to Z = 1$. Lastly, we verify the actual operation by $X \to Z = Y$ coupled with $X \wedge W = 0$. □

*Proof for $op \supseteq \{/, inc, Bool\}$.* As before, we work from Lemma 9 and only need to complete two details.

1. How does one unpack $K$ general integers from a single $\alpha$?

2. How does one simulate left-shifting using $op$?

Extracting multiple integers from $\alpha$ is done in this case by storing in $\alpha$ the contents of the vector $[2^u - 1, 0, a_1, \ldots, a_K]$, where $u$ is the width of the vector. Extracting $2^u - 1$ is done as before, and the rest of the elements can be retrieved by repeatedly dividing by $2^u$, and then ultimately taking the bitwise intersection with $2^u - 1$.

Next, we wish to simulate $X \Leftarrow Y \leftarrow Z$, which, as before, is done by verifying $X = Y \leftarrow Z$. Specifically, note that $Z$ is always either a constant or one of $w_1$ and $w_3$, and that this is the only context in which either $w_1$ or $w_3$ are ever used. Instead of storing these values directly, we therefore store $Z_1 = 2^{w_1} - 1$ and $Z_3 = 2^{w_3} - 1$, instead.

Ascertain that $Z_1$ and $Z_3$ are valid can be performed by $\mathrm{inc}(Z_i) \wedge Z_i = 0$. Following the transformation from $w_i$ to $Z_i$, the left shifts become multiplications: $X = Y \times \mathrm{inc}(Z_i)$. We verify them by $X \wedge Z_i = 0$ and $Y = X/\mathrm{inc}(Z_i)$, where the first comparison ascertains that the second comparison involves an exact division (utilizing the fact that $\mathrm{inc}(Z_i)$ is a power of two). □

*Proof for $op \supseteq \{\times, inc, Bool\}$.* We use the same $\alpha$ as in the proof for $op \supseteq \{/, inc, Bool\}$, noting that $X = Y \leftarrow w_i$ can be verified now simply by $X = Y \times Z_i$. This leaves us with the need to extract multiple integers, $a_1, \ldots, a_K$ from a single $\alpha$, which is done by

$$a_i \times 2^{u(i+1)} \Leftarrow ((2^u - 1) \overbrace{\times 2^u \times \cdots \times 2^u}^{i+1 \text{ times}}) \wedge \alpha.$$

Notably, the $a_i$ themselves are not extracted, but rather a shifted version of them. The reason that this partial extraction suffices is because in all operations, $X \times Z$ will be offset by a known power of $2^u$ compared to the offset with which we calculate $Y$, and simply multiplying one side by $2^u$ the appropriate (and bounded) number of times will serve to prove equality. □

# F Some useful basic operations

We first note that "$\dotdiv$" can always be implemented as an O(1) operation when "+" and Boolean functions are available. This can be done as follows.

$$a \dotdiv b = \begin{cases} 0 & \text{if } \text{SET}(a) \vee \text{SET}(b) \neq \text{SET}(a), \\ 0 & \text{if } \text{SET}(a) = \text{SET}(b) \\ & \text{and } (a + \neg b) \wedge (\text{SET}(a) + 1) = 0, \\ (a + \neg(b + \text{SET}(a))) \wedge \text{SET}(a) & \text{else,} \end{cases}$$

where $\text{SET}(a) \stackrel{\text{def}}{=} a \vee \neg a$. As such, "$\dotdiv$" should always appear as an optional operation wherever addition and Boolean functions are part of a RAM's basic operation set.

Our definition of $O_{a,m}^n$ was given originally as

$$O_{a,m}^n \stackrel{\text{def}}{=} ((a \leftarrow nm) \dotdiv a)/((1 \leftarrow m) \dotdiv 1).$$

With the optional "$\dotdiv$", this can be implemented directly using the operations available in Theorem 3, except for the multiplication "$nm$". However, if $n$ is known to equal $2^T$, this product can be calculated by means of a left shift: $nm = m \leftarrow T$.

We now construct a function, GT, that takes two encoded vectors, $V_1$ and $V_2$, each of width $m$ and length $n$, that returns another vector, of the same length and width, that has a 1 in every position where the decoded $V_1$ is greater than the decoded $V_2$, and 0 otherwise. This is described as Algorithm 3.

---

**Algorithm 3** $\text{GT}(m, V_1, V_2, n)$

1: $\text{MSB} \Leftarrow O_{1 \leftarrow (m \dotdiv 1), m}^n$
2: $\text{Mask} \Leftarrow O_{\neg(1 \leftarrow (m \dotdiv 1)), m}^n$
3: $\text{MSB}_1 \Leftarrow V_1 \wedge \text{MSB}$
4: $\text{MSB}_2 \Leftarrow V_2 \wedge \text{MSB}$
5: $\text{Mask}_1 \Leftarrow V_1 \wedge \text{Mask}$
6: $\text{Mask}_2 \Leftarrow V_2 \wedge \text{Mask}$
7: $\text{CarryToMSB} \Leftarrow ((\text{Mask}_1 + \text{Mask}) \dotdiv \text{Mask}_2) \wedge \text{MSB}$
8: $\text{Carry} \Leftarrow (\text{CarryToMSB} \wedge \text{MSB}_1) \vee ((\text{CarryToMSB} \vee \text{MSB}_1) \text{ clear } \text{MSB}_2)$
9: **return** $\text{Carry} \rightarrow (m \dotdiv 1)$

---

The algorithm essentially calculates the overflow bit in each element during the subtraction operation $V_1 - V_2$, but uses special handling for the MSB so as to ensure that the overflow from no element affects the result in any other element.

The operations used are all available to Theorem 3 directly, except for the right shift on the last line, which is accomplished by means of Lemma 8. (Nowhere in the proof of Theorem 3 is division applied to operands that were calculated by use of right shifts. This being the case, Lemma 8 continues to be applicable despite the larger operation set available to Theorem 3.)

The function GT now allows us to construct a test for equality as follows.

$$\mathrm{EQ}(m, V_1, V_2, n) = O^n_{1,m} \mathbin{\dot{-}} \mathrm{GT}(m, V_1, V_2, n) \mathbin{\dot{-}} \mathrm{GT}(m, V_2, V_1, n).$$